\documentclass[a4paper]{jpconf}
\usepackage{graphicx}
\usepackage{amsmath, amsthm, amssymb}
\begin{document}
\title{Kovacs effect in  solvable model glasses}
\author{Gerardo Aquino$^{1}$, Luca Leuzzi$^{2}$ and  Theo M. Nieuwenhuizen$^{1}$}
\address{$^{1}$ Institute for Theoretical Physics, University of Amsterdam,
Valckenierstraat 65, 1018 XE Amsterdam, The Netherlands}
\address{$^{2}$Department of Physics, University of Rome ``La Sapienza'', Piazzale A.Moro 2, 00185 Roma, Italy  and  Institute of  Complex Systems
(ISC)-CNR, via dei Taurini 19, 00185 Roma, Italy}
\ead{gaquino@science.uva.nl}
\begin{abstract}
The Kovacs protocol, based on the temperature shift experiment originally
conceived  by A.J. Kovacs and applied on  glassy polymers \cite{kovacs},
 is implemented in an exactly solvable   model with facilitated dynamics.
This model is based on
interacting fast and slow modes represented respectively by spherical spins and harmonic
oscillator variables. Due to this fundamental property and to slow dynamics, the  model reproduces
the characteristic  non-monotonic evolution  known as the ``Kovacs effect'', observed
 in polymers,  
 spin glasses,
 in granular materials
and models of molecular liquids,
 when similar experimental protocols are implemented.
\end{abstract}



\section{Introduction}
Glassy systems,  being in an out-equilibrium condition, have properties
 which  depend on their history. This is the 'memory' of glasses.
This property can manifest itself in   striking ways, when specially devised  experiments are made. One example  is given by negative
temperature cycles in spin glasses where the ac susceptibility, depending on both frequency and the age of the system, recovers the exact value it had before the negative temperature jump.
 A  memory effect which  shows up in a  one-time observable,  when a specific experimental protocol is implemented, is the so called ``Kovacs effect'' \cite{kovacs}.  This effect has been the subject of a variety of recent investigations \cite{berthier, buhot, bertin, cuglia, mossa, sellitto}.   The characteristic
non-monotonic evolution of the observable under examination (the volume in the original Kovacs' experiment), with the other thermodynamic variables held constant, shows clearly that a
 non-equilibrium state of the system cannot be fully characterized only by the (time-dependent) values
 of  thermodynamic variables, but that further
inner variables are needed to give a full description of the non-equilibrium state of the
system. The memory  in this case consists in these internal variables keeping track
of the  history of the system.\\
\indent  The purpose of this paper is to use a specific model for fragile
glass  to implement the protocol and get some insight into the Kovacs effect. We show
that in spite of its simplicity, this model captures the phenomenology of the Kovacs effect 
 and allows in specific regimes to obtain analytical expressions for the evolution of the variable of interest. 
 \indent This paper is organized as follows: in Section II we review the experimental protocol generating the effect, in Sections III and IV we introduce our model and
use it to implement the protocol,in section V we draw out of this model some analytical results
 with final conclusions.
An appendix collects all terms and coefficients
employed in the main text.

\section{Kovacs protocol}

The experimental protocol, as originally designed by A. J. Kovacs in the '60s \cite{kovacs}, consists of three main steps:
\begin{description}
\item[1$^{st}$ step]
 The system is equilibrated at a given high temperature $T_i$. 
\item[2$^{nd}$ step] 
 At time $t=0$ the system is  quenched to a lower temperature $T_l$, close to or  below
the glass transition temperature,  and it is let to evolve a period $t_a$. One  then follows the evolution of the
  the proper thermodynamic variable
 (in the original Kovacs experiment this was the volume $V(t)$ of a sample of polyvinyl acetate, in our model
  it will be the ``magnetization''  $M_1(t)$).
\item[3$^{rd}$ step] 
 After the time $t_a$, the volume, or other corresponding observable, has reached a value  equal,  by definition of $t_a$, to the
equilibrium value corresponding  to an intermediate temperature $T_f$ ($T_l<T_f<T_i$), i.e. such that  $V_{T_l}(t_a)\equiv V^{eq}_{T_f}$.
 At this time, the bath temperature  is switched  to  $T_f$.

 The pressure (or corresponding variable) is kept constant throughout the whole experiment.
 \end{description}

 Naively one would expect the observable under consideration, after the third step, to remain constant   since it already  has (at time $ t=t_{a}^{+})$ its equilibrium value. But the system  has not
  equilibrated yet and so the observable  goes through a non monotonic evolution before
relaxing back to its equilibrium value, showing a characteristic hump whose maximum increases
with the magnitude of the final jump of temperature $T_f-T_l$ and occurs at a time which decreases
with increasing  $T_f-T_l$.

 We want  to implement this protocol on a model for both strong and fragile glass first
 introduced in \cite{lucateo1}: the Harmonic Oscillators-Spherical Spins model (HOSS). This model is based on interacting fast and slow modes,
 this property turns out to be necessary for the memory effect, object of this paper, to occur.
\newcommand{\pz}{\partial}
\section{The Harmonic Oscillator-Spherical Spin Model}

 The HOSS model contains a set of $N$ spins $S_i$ locally coupled to a set of $N$ harmonic
 oscillator  $x_i$ according to the following hamiltonian:
\begin{equation}
{ \cal{ H}}=\sum_{i=1}^N (\frac{K}{2} x_{i}^2 -H x_i -J x_i S_i -L S_i)
\end{equation}
The spins have no fixed length but satisfy the spherical constraint: $\sum_{i=1}^N S_i^2=N$.
 The spin variables are assumed to relax on a much shorter time scale than the harmonic
oscillator variables, so the oscillator variables are the slow modes and on their dynamical
evolution the fast spin modes act just as noise. One can then integrate out the spin
variables to obtain the following effective Hamiltonian for the oscillators (for details see
\cite{lucateo1},
  explicit expressions of undefined  terms appearing in all  equations hereafter are reported in  the
Appendix):
\begin{eqnarray}\label{heff}
\frac{{\cal{H}}_{\it eff}(\{x_i\})}{N}=\frac{K}{2} M_2 -H M_1 -w_T(M_1,M_2) +\frac{T}{2}\log \left(\frac{w_{T}(M_1,M_2)+\frac{T}{2}}{\frac{T}{2}}\right)
\end{eqnarray}
which depends on the temperature and on the  first and second moment of the oscillator variables, namely:
\begin{eqnarray}
M_1=\frac{1}{N} \sum_{i=1}^N x_i \; \; \;, \; \; \; M_2=\frac{1}{N} \sum_{i=1}^N x_i^2
\end{eqnarray}
These variables encode the dynamics of the system which is
 implemented through a Monte Carlo
 parallel update of the oscillator variables:
\begin{equation}\label{update}
  x_i \to x_i +r_i/\sqrt{N}
\end{equation}
The variables $r_i$ are normally distributed with zero mean value and variance $\sigma^2$.
The update is accepted according to the Metropolis acceptance rule applied to the variation $\delta \epsilon$
of the energy   of the oscillator variables, which is determined by  ${\cal{H}}_{\it eff}$ and, in the limit of large $N$, is given by:
\begin{equation}\label{increment}
\frac{\delta \epsilon}{N} =\frac{K_T(M_1,M_2)}{2}\delta M_2 -H_T(M_1,M_2)\delta M_1 .
\end{equation}
 This simple model turns out to have a slow dynamics and can be solved analytically.

Following \cite{lucateo1} one can derive the dynamical equations for $M_1$ and $M_2$
\begin{eqnarray}\label{evolution}
\dot{M}_1&=&\left[\frac{H_T(M_1,M_2)}{K_T(M_1,M_2)}-M_1\right]f_T(M_1,M_2)\\
\nonumber \dot{M}_2 &=&\frac{2}{K_T(M_1,M_2)}\left[I_{T}(M_1,M_2)+H_T(M_1,M_2)\dot{M}_1\right]
\end{eqnarray}
The stationary solutions of these equations coincide with the saddle point of the partition
function of the whole system at equilibrium at temperature $T$  and are given by:
 \begin{eqnarray}\label{equilibrium}
      \bar{M}_1&=&\frac{H_T(\bar{M}_1,\bar{M}_2)}{K_T(\bar{M}_1,\bar{M}_2)}=\frac{\bar{H}_T}{\bar{K}_T}\\
      \nonumber \bar{M}_2-\bar{M}_1^2&=&\frac{T}{K_T(\bar{M}_1,\bar{M}_2)}=\frac{T}{\bar{K}_T}
       \end{eqnarray}
with barred variables from now on indicating their equilibrium values.

\subsection{Strong and Fragile Glasses with the HOSS model}
 In spite of its simplicity, the HOSS model
 allows  to describe both
strong and fragile glasses, characterized respectively by an Arrhenius or a Vogel-Fulcher
law in the relaxation time.
 The following constraint on the configurations space is applied:
\begin{equation}
\mu_2=M_2-M_1^2-M_0 \geq 0
\end{equation}
When $M_0=0$ there exists a single global minimum in the configurations space of the
oscillators, therefore the role of the constraint with $M_0>0$  is to avoid the existence of
a ``crystalline state'' and to introduce a finite transition temperature. The stationary
solutions for the dynamics  with this constraint  are given by:
 \begin{eqnarray}\label{equilibrium2}
      \bar{M}_1&=&\frac{H_T(\bar{M}_1,\bar{M}_2)}{K_T(\bar{M}_1,\bar{M}_2)}=\frac{\bar{H}_T}{\bar{K}_T}\\
      \nonumber \bar{M}_2-\bar{M}_1^2&=& \left\{
               \rule{0 cm}{0.75 cm}
                  \begin{array}{c}
                  \frac{T}{K_T(\bar{M}_1,\bar{M}_2)}=\frac{T}{\bar{K}_T} \; \;\;\; T >T_k\\
                    \;  \\
                   M_0                \;\;\;\;\;\;\;\;   \;\;\;\;\;\;\;\;\;  \;\;\;\;\;\;\;\;        T\leq T_k
                    \end{array}
                   \right.
       \end{eqnarray}
   The temperature $T_k$ is determined by the further condition:
  \begin{equation}\label{kauz}
    T_k=M_0  \; K_{T_k}(\bar{M}_1^{T_k},\bar{M}_2^{T_k})=M_0 \; \bar{K}_{T_k}.
  \end{equation}
This is the highest temperature at which the constraint is fulfilled, for smaller  temperatures the system relaxes to equilibrium configurations which fulfill the constraint.
For $T>T_k$  therefore the dynamics is not affected by the constraint.
For $T \leq T_k$ the system eventually reaches  a configuration which fulfills the constraint, when this happens  it gets trapped for ever in such a configuration.
 This is equivalent to having a ``Kauzmann-like'' transition, occurring at $T=T_k$ with  vanishing configuration entropy, meaning the system gets stuck forever in one single configuration fulfilling the constraint (see also: \cite{teocond}).

When there is no constraint, i.e. when $M_0=0$, then  $T_k=0$, if the Monte Carlo updates are done with Gaussian variables
with constant variance $\sigma^2$, this model is characterized by an Arrhenius  relaxation
				 law:
\begin{equation}\label{Arrhenius}
  \tau_{eq}\sim e^{\frac{A_s}{T}}
  \end{equation}
  in so resembling the relaxation properties of strong glasses.

 The HOSS model with constraint strictly positive ($ M_0 >0$) can  easily be extended to describe fragile glasses by
  further introducing in the variance of the Monte Carlo update, the following
       dependence on the dynamics:
\begin{equation}\label{deltadep}
 \sigma^2=8(M_2-M_1^2)(M_2-M_1^2-M_0)^{-\gamma}
\end{equation}
In this case the relaxation time turns out to follow the generalized Vogel-Fulcher law:
\begin{equation}\label{V-K}
 \tau_{eq}\sim e^{A_k^{\gamma}/(T-T_k)^{\gamma}}
\end{equation}
The parameter $\gamma$ is introduced to make the best Vogel-Fulcher type fit for the relaxation
time in experiments, making this model valid for a wide range of fragile glasses.
When the temperature approaches the value $T_k$  defined by (\ref{kauz}), from above, the system relaxes towards configurations close to the ones fulfilling the constraint.
The variance $\sigma^2$ then tends to diverge,
the updates become large and so unfavorable, meaning
that almost every update of the oscillator variables is refused.
This produces the diverging relaxation time  following the Vogel-Fulcher law of Eq. (\ref{V-K}).

\section{Kovacs effect in the {\sl{HOSS}} model}
\begin{figure}[h]
\begin{minipage}{18pc}
\includegraphics[width=18pc]{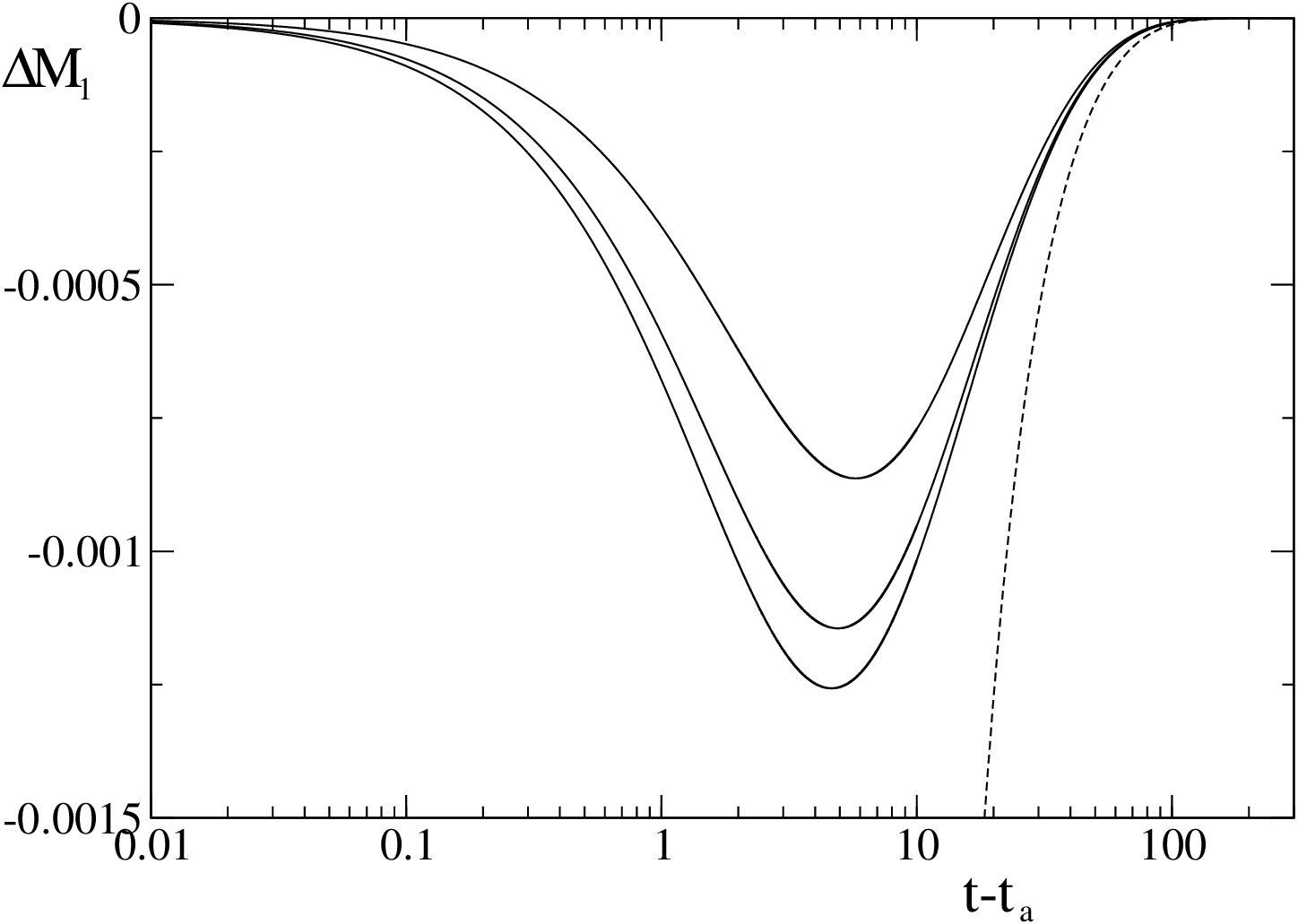}
\caption{\label{fig2}   {\it Fragile glass with $\gamma$=$1$} The Kovacs protocol is implemented with a quench  of the system from
temperature $T_i$=$10$ to $T_l$, and final switch (at $t$=$t_a^+$) to the intermediate temperature $T_f$=$4.3$. The continuous lines, starting from the lowest, refer to $T_l$=$4.005, 4.05, 4.15$, the dashed line refers to condition $T_l=T_f$, ( simple aging with no final temperature shift).}
\end{minipage}\hspace{2pc}%
\begin{minipage}{18pc}
\includegraphics[width=18pc]{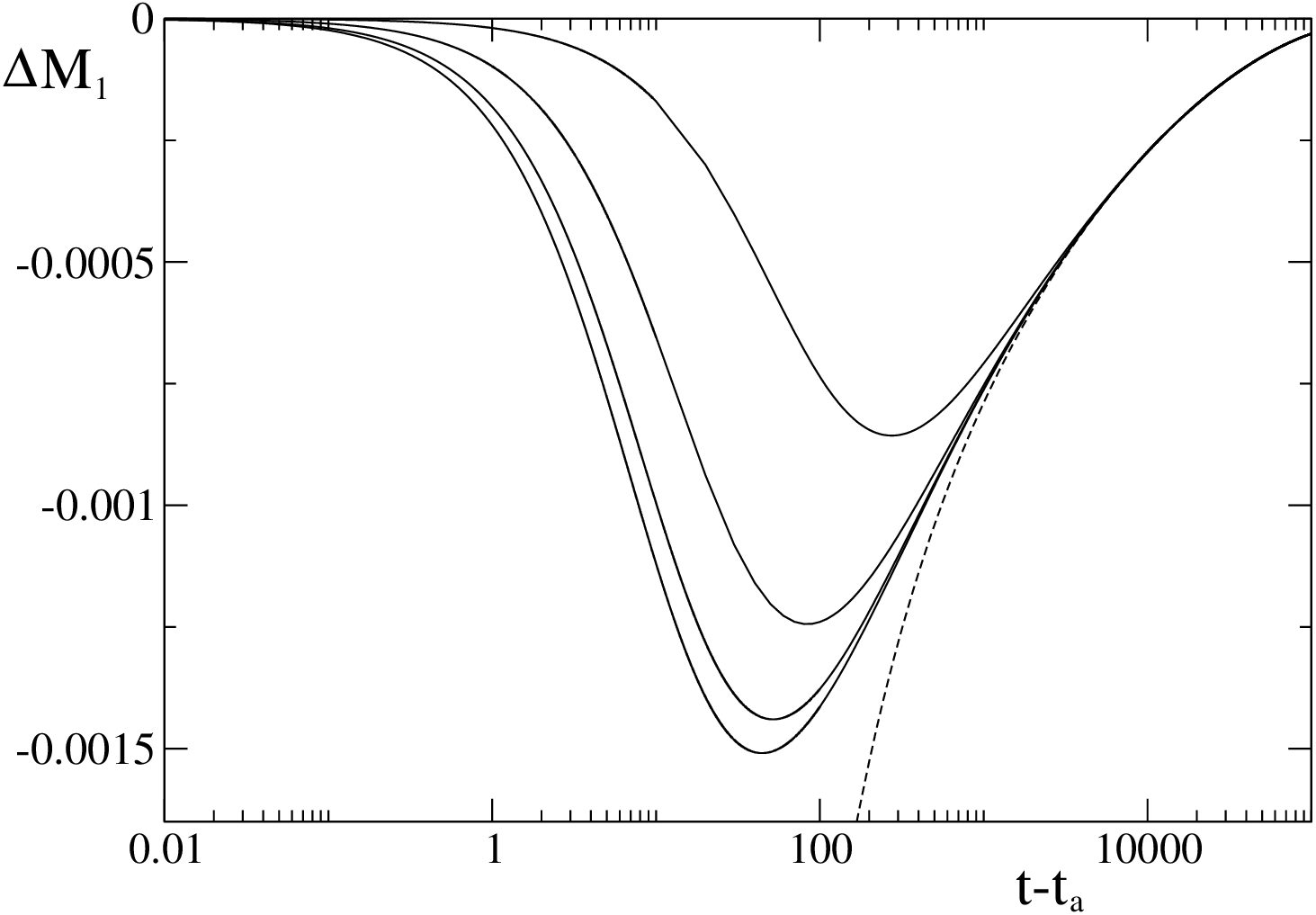}
\caption{\label{fig3} {\it Fragile glass with $\gamma$=$2$}. The Kovacs protocol is  implemented with a quench of the system
  from temperature $T_i$=$10$ to $T_l$, and final switch (at $t$=$t_a^+$) to the intermediate
   temperature $T_f$=$4.3$. The continuous lines, starting from the lowest, refer to $T_l$=$4.005, 4.05, 4.15, 4.25$, the dashed line refers to condition $T_l$=$T_f$ (simple aging with no final temperature shift).}
\end{minipage} 
\end{figure}

We implement the Kovacs protocol in the model above introduced for a fragile glass. The
system is prepared at a temperature $T_i$ and quenched to a region of temperature close to
the $T_k$, i.e. $T_l \gtrsim T_k$.  Solving numerically Eqs. (\ref{evolution}) we determine
the evolution of the system in both step 2 and 3 of the protocol. In step 2 the time $t_a$,
at which $M_1^{T_l}(t_a)=\bar{M}_1^{T_f}$, is calculated so that:
\begin{eqnarray}\label{continuity}
{M}_1^{T_f}(t_a^+)&=&\bar{M}_1^{T_f}\\
\nonumber {M}_2^{T_f}(t_a^+)&=&{M}_2^{T_l}(t_a)
\end{eqnarray}
The evolution of the  fractional "magnetization":
\begin{equation}
\vspace{-0.1cm}
   \Delta M_1(t)=\frac{M_1(t)-\bar{M}_1^{T_f}}{\bar{M}_1^{T_f}}
\vspace{-0.1 cm}
\end{equation}
 after step 3 ($t>t_a$) for different values of $T_l$ is reported
in Figs. \ref{fig2} and \ref{fig3} respectively for $\gamma=1$ and $\gamma=2$.  
The magnetic field $H$ is kept constant at the value $H=0.1$.  
In all the implementations of the protocol we   use
the  values $J$=$K$=$1$, $L$=$0.1$ and $M_0$=$5$ for the other  parameters of the model.
This choice  for the parameters  and the value $H$=$0.1$ for the magnetic field
 fix (through  Eqs. (\ref{equilibrium2}) and (\ref{kauz})) the Kauzmann temperature at  the value $T_k$=$4.00248$. 

 Since the equilibrium value of $M_1$ decreases with increasing temperature (as opposed to
what happens for instance with the volume) we observe a reversed 'Kovacs hump'. The curves
keep the same properties typical of the Kovacs effect, the minima occur at a time which
decreases and have a depth that increases with  increasing magnitude of the final switch of
temperature. As expected, since increasing $\gamma $ corresponds to further slowing the dynamics, the effect shows on a longer time scale in the case $\gamma=2$ as compared to $\gamma=1$.

Actually, since in the last step of the protocol: $M_1(t=t_a)=\bar{M}_1^{T_f}$ and
 $f_{T_f}(M_1,M_2)$ is always positive, from  the first of  Eqs. (\ref{evolution})
 one soon realizes that the hump
 for this model can be either positive or negative, depending on the sign of the term:
\begin{equation}\label{humpdefine}
\frac{H_{T_f}(\bar{M}_1^{T_f},M_2)}{K_{T_f}(\bar{M}_1^{T_f},M_2)}-\bar{M}_1^{T_f}
\end{equation}
at $t=t_a^{+}$.
This term is zero when $M_1=\bar{M}_1^{T_f}, M_2=\bar{M}_2^{T_f}$, so one would expect
$M_2(t=t_a^+)=\bar{M}_2^{T_f}$ to be the border value determining the positivity or negativity of the
hump.
Since $H_{T_f}(\bar{M}_1^{T_f},M_2)$ decreases with increasing $M_2$ while
 $K_{T_f}(\bar{M}_1^{T_f},M_2)$ increases, it follows that the condition 
for a positive hump is:
\begin{equation}
M_2(t=t_a^+)< \bar{M}_2^{T_f}
\end{equation}
 For  shifts of temperature in a wide range close  to the transition temperature
$T_k$, where the dynamics is slower and the effect is expected to show up significantly on a
long time scale, the condition $M_2(t=t_a)>\bar{M}_2^{T_f} $ is always fulfilled and
therefore a negative hump is expected.


\section{Analytical solution in the  long-time regime}
 In the previous Section   we have shown, through a numerical solution of the
 dynamics, that the HOSS model reproduces the phenomenology of the Kovacs effect,
 showing the same qualitative properties of the Kovacs hump as obtained in experiments (see for ex. \cite{kovacs,josserand}),
  in  some other models with  facilitated or kinetically constrained dynamics \cite{sellitto, buhot} and in  other different  models \cite{berthier, bertin, cuglia,mossa}.
 In this section we show that, by carefully choosing the working conditions in which the
protocol is implemented, our model provides with an analytical solution for the evolution of
the variable of interest.
\subsection{Auxiliary variables}
In order to ease calculations, as done in \cite{teocond, lucateo1} it is convenient to introduce the following
variables:
\begin{eqnarray}\label{muvariables}
&&\mu_1=\frac{H_T(M_1,M_2)}{K_T(M_1,M_2)}-M_1\\
\nonumber &&\mu_2=M_2-M_1^2-M_0
\end{eqnarray}
for which the dynamical equations read:
\begin{eqnarray}\label{mu-evolution}
\dot{\mu}_1 &=& -J Q_T(M_1,M_2)I_T(M_1,M_2) -(1+ D Q_T(M_1,M_2))\mu_1 f_T(M_1,M_2)\\
\nonumber \dot{\mu}_2&=&\frac{2I_{T}(M_1,M_2)}{K_T(M_1,M_2)}+2 \mu_1^2 f_T(M_1,M_2)
\end{eqnarray}

 We will choose to implement steps 2 and 3 of the protocol in a range of temperature very
close to the Kauzmann temperature $T_k$. As exhaustively shown in \cite{lucateo1, luca} in
the long time regime the variable $\mu_2(t)$ decays logarithmically   to its equilibrium
value which is  small for $T \sim T_k$.
 So, if $t_a$ is very large, the value of the variable
$\mu_2(t)$, which is continuous at the jump,  will be small enough to fulfill the condition
for which the following equation is shown to be valid \cite{lucateo1}:
\begin{eqnarray}\label{assaver}
 \frac{d \mu_1}{d(\delta \mu_2)}= (1+Q_T(M_1,M_2) D)\frac{(\bar{\mu}_2+\delta \mu_2)^{-\gamma}}{\delta \mu_2}\mu_1 - \frac{J Q_T(M_1,M_2) T}{2(M_0+\bar{\mu}_2)}
\end{eqnarray}
where now the variable $\delta \mu_2(t)=\mu_2(t)-\bar{\mu}_2 $ is used and barred variables
always refer to equilibrium condition. Of course  choosing $T_l$ close to $T_k$ and waiting a
long time $t_a$ so that the system approaches equilibrium, allows only small temperature
shifts for the final step of the protocol, meaning that also $T_f$ will be close to $T_k$.
 All the coefficients   which appear in
equation (\ref{assaver}) (see Appendix for complete expressions) in the regime chosen, can be
assumed constant and equal to their equilibrium values with a very good approximation. The
equation can then be easily integrated to give:
\begin{equation}\label{assaver0}
\nonumber  \mu_1(\delta \mu_2)= \exp\left[-\frac{\,  _2 F_1(\gamma,\gamma,\gamma+1,-\frac{\bar{\mu}_2}{\delta \mu_2})}{\gamma (\delta \mu_2)^{\gamma}/A_Q}\right]
\left(\mu_1^+ B_Q^{\gamma} -C_Q \int_{\delta \mu_2^+}^{\delta \mu_2} dz \exp\left[\frac{ \, _2 F_1(\gamma,\gamma,\gamma+1,-\frac{\bar{\mu}_2}{z})}{\gamma z^{\gamma}/A_Q}\right]\right)
\end{equation}
where the superscript $^+$ indicates $t=t_a^+$ and $_2 F_1$ the hypergeometric function. This expression simplifies in cases $
\gamma=1,\; 3/2,\; 2$. All these solutions and relative coefficients are reported in the
appendix, here we limit ourselves to the case $\gamma=1$ which corresponds to ordinary
Vogel-Fulcher relaxation law. In this case the solution is:
\begin{eqnarray}\label{assaver1}
 \mu_1(t)=\left( \frac{\delta \mu_2(t)}{\delta \mu_2(t) +\bar{\mu}_2}
 \right)^{\frac{A_Q}{\bar{\mu}_2}} \left[\mu_1^+
 \left( \frac{\delta \mu_2^+ +\bar{\mu}_2}{\delta \mu_2^+} \right)^{\frac{A_Q}{\bar{\mu}_2}}   -C_Q \int_{\delta \mu_2^+}^{\delta \mu_2(t)} dz\left( \frac{z}{z
+\bar{\mu}_2}\right)^{-\frac{A_Q}{\bar{\mu}_2}} \right]
\end{eqnarray}
where:
\begin{eqnarray}
\nonumber \int_a^b dz\left( \frac{z}{z+ \eta }\right)^{\alpha}= 
\frac{x^{\alpha+1} \;  _2 F_1(1-\alpha,-\alpha, 2-\alpha,-\frac{x}{\eta})}{\eta^{\alpha}(1+\alpha)}  |_{x=a}^{x=b}
\end{eqnarray}
One can then expand the variable of interest $M_1(t)$ in terms of $\mu_1$ and
$\delta \mu_2$ and obtain the following  expression for the Kovacs curves:
\begin{eqnarray}\label{m1approx}
 \delta M_1(t)=A_{T_f}^1(\bar{M}_1^{T_f},\bar{M}_2^{T_f}) (\mu_1(t) -\mu_1^+) +A_{T_f}^2(\bar{M}_1^{T_f},\bar{M}_2^{T_f})(\delta \mu_2(t)-\delta\mu_2^+)         
\end{eqnarray}
where the coefficients are approximately constant in the regime chosen and can be evaluated at equilibrium.
\subsection{Short and intermediate $t-t_a$}

For  small $t-t_a$,
 a linear
 approximation for the variable $\delta \mu_2$, with slope given by the second equation of the
set (\ref{mu-evolution}) evaluated at $t=t_a^+$, turns out to be very good.
Inserting this expression in Eq. (\ref{assaver1}) to get $\mu_1(t)$ and then in Eq.(\ref{m1approx})
 a good approximation of the first part of the hump  for small and
intermediate $t-t_a$ is obtained, as shown in Fig. \ref{fig5}.
\subsection{Intermediate and long $t-t_a$ }
\begin{figure}
\begin{center}
\includegraphics[width=8.0cm, height=5.7 cm,angle=0]{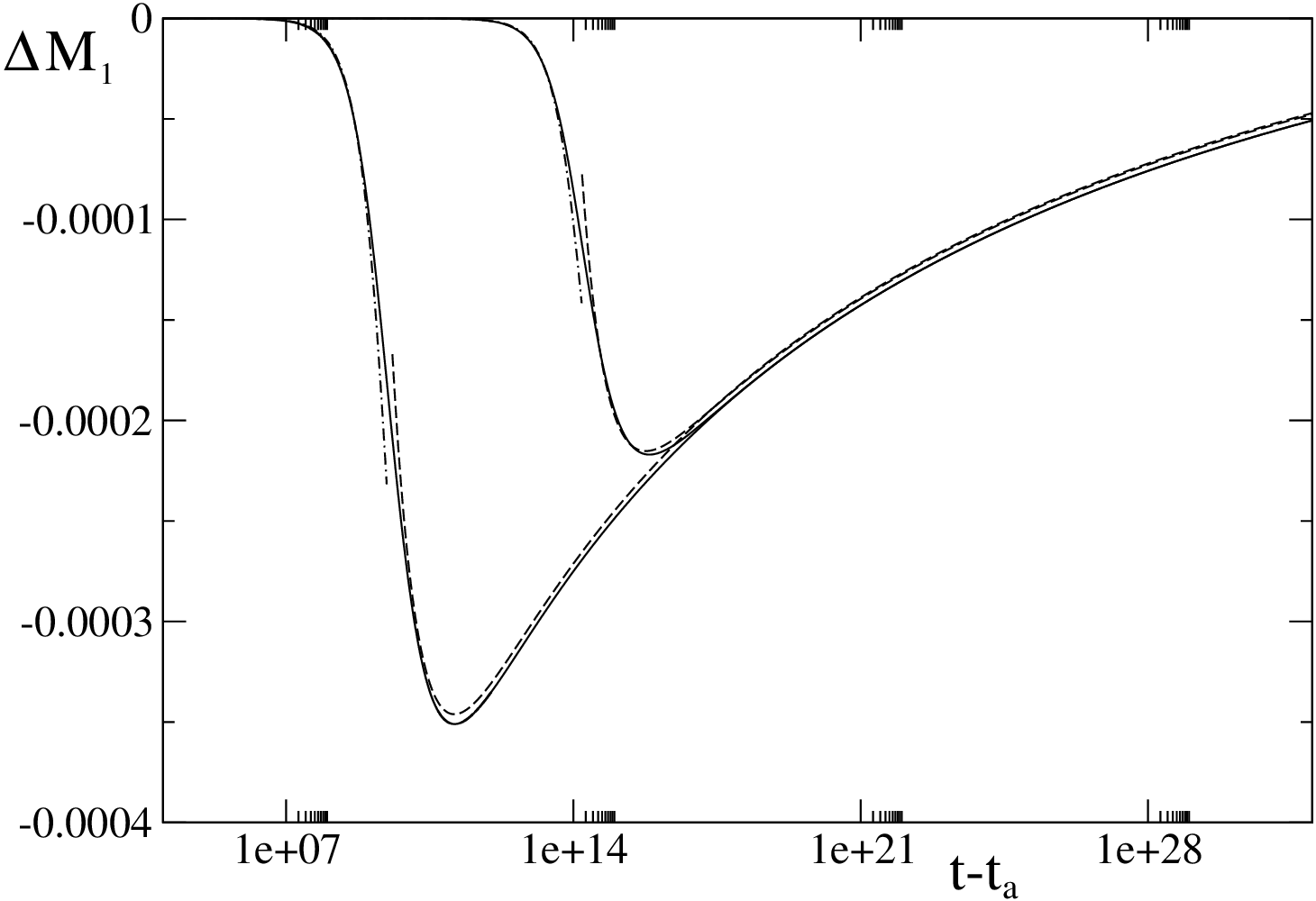}
\end{center}
\caption{\label{fig5} Comparison between numerical solution (continuous lines) for the Kovacs' curves   and the  approximate analytical solution at short-intermediate (dot-dashed line) and intermediate-long time (dashed line). The protocol is implemented between $T_i=10$ and $T_f=4.018$. The curves starting from the lowest refer to $T_l=4.005$, $4.008$. ($H=0.1$, $T_k=4.00248$)}
\end{figure}

 When $t-t_a$ is very large, we can use  Eq. (\ref{assaver1}) and the pre-asymptotic approximation for $\mu_2(t)$
(see: \cite{lucateo1})
\begin{equation}
\mu_2(t)=\left(\log(t/t_0)+ \frac{1}{2} \log(\log(t/t_0))\right)^{-1/\gamma}
\end{equation}
Inserting this expression in Eq. (\ref{assaver1}) to get $\mu_1(t)$ and then in Eq.(\ref{m1approx}), a good approximation for the hump and the tail of the Kovacs curves is obtained.
In Fig. \ref{fig5} we show the agreement between the analytical expression so obtained 
  and the numerical solution.


\section{Conclusion}

We have shown that a simple mode with constrained dynamics like the HOSS model, is rich enough to reproduce the Kovacs memory effect, even  allowing to obtain analytical expression for the Kovacs
hump in a long time regime.
 The Kovacs effect is observed in many experiments and models, showing common qualitative
properties which we have found to be shared also by the model analyzed in this paper. The
quantitative properties depend on the particular system or model analyzed.

 As far as it concerns
the HOSS model, it turns out that for the slow modes, i.e.  the oscillator variables,
 fixing the overall  average value, the magnetization $M_1$, does not prevent the existence of 
 memory encoded in the variable $M_2$, which keeps track of the  history of the system.
 The equilibrium value of $M_2$
  increases with temperature while the equilibrium value of $M_1$ decreases with increasing
  temperature. Therefore after the final switch of temperature, since $M_2(t_a)>\bar{M}_2^{T_f} $,
the variable $M_2$ has a value corresponding to an equilibrium condition at a higher temperature (memory of the initial
state at temperature $T_i$) so driving the system  towards a condition corresponding to a
higher temperature,  i.e. smaller values of  $M_1$, determining the hump.

 It is important to stress that a fundamental
ingredient in the HOSS model, besides the slow dynamics which originates 
from the Monte Carlo parallel update,  is  the interaction between slow and fast modes. Due to this
interaction the equilibrium configurations  of the oscillator variables at a given
temperature are determined by both $M_2$ and $M_1$, the first and second moment of their
distribution, whose dynamical evolution is interdependent.  When such interaction is turned
off (by setting $J=0$ ) essentially only one variable is sufficient to describe the equilibrium configurations and the dynamics of the system, and the memory effect is lost.
 In this respect this model constitutes an improvement to the so-called oscillator model
 \cite{bonilla} within which such memory effect cannot be reproduced.
In the present model one can also study temperature cycle experiments of the
type carried out in spin glasses (see. \cite{conf}), leaving room
for further research. \\
More details on this subject can be found in Ref. \cite{condaq}.\\
 
 G.A. and L.L. gratefully  acknowledge the European network  DYGLAGEMEM for financial support.
\appendix
\section{}
In this Appendix we report all the explicit expressions for terms appearing in the text. In
Eqs. (\ref{heff}), (\ref{increment}) and (\ref{evolution})  we have:
\begin{eqnarray}
 \nonumber &&w_T(M_1,M_2)=\sqrt{J^2 M_2 +2 J L M_1+L^2+T^2/4}\\
  \nonumber&&K_T(M_1,M_2)=K-\frac{J^2}{w_T(M_1,M_2)+T/2}\\
\nonumber&&H_T(M_1,M_2)=H+\frac{J L}{w_T(M_1,M_2)+T/2}\\
&&\nonumber f_T(M_1,M_2)=\frac{\sigma^2 K_T(M_1,M_2)}{2 T}Erfc\left[\tilde{\alpha}_T(M_1,M_2)\right]\cdot \exp\left[\tilde{\alpha}_T^2(M_1,M_2)-\alpha_T^2(M_1,M_2) \right]\\
&&\nonumber
I_T(M_1,M_2)=\frac{\sigma^2 K_T(M_1,M_2)}{4}  Erfc\left[\alpha_T(M_1,M_2)\right]+ \left(\frac{T}{2} -K_T(M_1,M_2) \tilde{w}_T(M_1,M_2)\right)f_T(M_1,M_2)
\end{eqnarray}
where:
\begin{eqnarray}
\nonumber&&\tilde{w}_T(M_1,M_2)=M_2-M_1^2+(\frac{H_T(M_1,M_2)}{K_T(M_1,M_2)}-M_1)^2\\
\nonumber&&\alpha_T(M_1,M_2)=\sqrt{\frac{\sigma^2}{8 \tilde{w}_T(M_1,M_2)}}\\
\nonumber &&\frac{\tilde{\alpha}_T(M_1,M_2)}{\alpha_T(M_1,M_2)}=\frac{2 K_T(M_1,M_2)
\tilde{w}_T(M_1,M_2)}{T}-1
\end{eqnarray}
In Eqs.  (\ref{Arrhenius}), (\ref{V-K}), (\ref{assaver}), (\ref{assaver0}), (\ref{assaver1})
and (\ref{m1approx}): \vspace{-0.0cm}
\begin{eqnarray}
\nonumber A_s&=&\frac{\sigma^2 \bar{K}_T}{8}\\
\nonumber D&=& J H + L K=J H_T +L K_T\\
\nonumber Q_T(M_1,M_2)&=&\frac{J^2 D}{K_T^3 w_T  (w_T +T/2)^2}\\
\nonumber P_T(M_1,M_2)&=&\frac{J^4 (M_2-M_1^2)}{2 K_T w_T  (w_T +T/2)^2}\\ 
\nonumber A_k&=&\frac{\bar{K}_{T_k}(K-\bar{K}_{T_k}) (1+ D\bar{Q}_{T_k}+\bar{P}_{T_k})}{(K-\bar{K}_{T_k})(1+D\bar{Q}_{T_k})-\bar{K}_{T_k}\bar{Q}_{T_k}}
\\
\nonumber  _2 F_1(a, b, c, z)&=&\frac{\Gamma(c)}{\Gamma(a) \Gamma(b)}\sum_{n=0}^{\infty} \frac{\Gamma(a+n) \Gamma(b+n) }{\Gamma(c+n)} \frac{z^n}{n!}\\
\nonumber A_Q&=&1+Q_{T_f}(\bar{M}_1^{T_f},\bar{M}_2^{T_f}) D= 1+\bar{Q}_{T_f}D\\
\nonumber B_Q^{\gamma}&=&\exp[A_Q\frac{\,  _2 F_1(\gamma,\gamma,\gamma+1,-\frac{\bar{\mu}_2}{\delta \mu_2^+})}{\gamma (\delta \mu_2^+)^{\gamma}}]\\
\nonumber C_Q&=&\frac{J Q_{T_f}(\bar{M}_1^{T_f},\bar{M}_2^{T_f})  T_f}{2(M_0+\bar{\mu}_2)}=\frac{J \bar{Q}_{T_f} T_f}{2(M_0+\bar{\mu}_2)}\\
\nonumber A^1_T(M_1,M_2)&=&\frac{(w_T +T/2)K_T}{M_1(J M_1 +L + (w_T+T/2)K_T)}\\
\nonumber A^2_T(M_1,M_2)&=& 2 M_1 A^1_T(M_1,M_2)\\
\nonumber t_0&=&\frac{\sqrt{\pi}}{8 \gamma}\frac{ 1+ D\bar{Q}_{T}}{1+D\bar{Q}_{T}+\bar{P}_{T}}
\end{eqnarray}

\end{document}